\documentclass[letterpaper,twocolumn,10pt]{article}
\usepackage{usenix,epsfig,endnotes}
\usepackage[T1]{fontenc}
\usepackage{graphicx}
\AtBeginDocument{%
  \providecommand\BibTeX{{%
    Bib\TeX}}}

\usepackage{amssymb}
\usepackage{amsmath,amsfonts}
\usepackage{algorithmic}
\usepackage{graphicx}
\usepackage{textcomp}
\usepackage{listings}
\usepackage{xcolor}

\usepackage{fontawesome5} 
\usepackage{graphicx}
\usepackage{booktabs}
\usepackage{array}
\usepackage{tcolorbox}
\usepackage{subcaption}
\usepackage{adjustbox}

\usepackage{multirow}
\usepackage{makecell}
\usepackage{array}
\usepackage{soul}

\usepackage{subcaption}
\usepackage{multirow} 
\usepackage{caption}  
\usepackage{colortbl} 
\usepackage[table]{xcolor} 

\usepackage{pifont}         
\usepackage{xcolor}         
\usepackage{tcolorbox}      
\usepackage{booktabs}       
\usepackage{array}          
\usepackage{mdframed}

\newcommand{\crossbox}{%
  \tcbox[colback=white, colframe=red, boxrule=0.4pt, arc=2pt, left=1pt, right=1pt, top=1pt, bottom=1pt, boxsep=0pt]{%
    \textcolor{red}{\ding{55}}%
  }%
}

\newcommand{\tickBoxGreen}{%
  \tcbox[colback=white, colframe=green!60!black,
    boxrule=0.5pt, arc=2pt, left=1pt, right=1pt, top=1pt, bottom=1pt, boxsep=0pt]{%
    \textcolor{green!60!black}{\ding{51}}%
  }%
}
\usepackage{xspace}
\newcommand{\name}{\texttt{MULTI-LLMSecCodeEval}\xspace}

\tcbset{
  /tcb/colback=gray!5,
  /tcb/colframe=black!50,
  /tcb/boxrule=0.5pt,
  /tcb/arc=2pt,
  /tcb/left=4pt,
  /tcb/right=4pt,
  /tcb/top=4pt,
  /tcb/bottom=4pt
}

\usepackage{pifont,xcolor}

\usepackage{listings}

\def\BibTeX{{\rm B\kern-.05em{\sc i\kern-.025em b}\kern-.08em
    T\kern-.1667em\lower.7ex\hbox{E}\kern-.125emX}}

\usepackage{tikz}
\usetikzlibrary{positioning, shapes, arrows.meta}

\definecolor{lightlight-gray}{gray}{0.97}
\global\mdfdefinestyle{rtboxstyle}{%
linecolor=black,%
leftmargin=0cm,rightmargin=0cm,linewidth=0.4pt,
roundcorner=2, skipabove=0.5em, innerleftmargin=5pt, innerrightmargin=5pt,
skipbelow=0pt,backgroundcolor=lightlight-gray
}

\newcommand{\rtbox}[1]{\begin{mdframed}[style=rtboxstyle]{{#1}}\end{mdframed}}

\definecolor{RED}{rgb}{1,0,0}
\newcommand{\updated}[1]{\textcolor{black}{#1}}

\begin{document}
\title{Does Teaming-Up LLMs Improve Secure Code Generation? A Comprehensive Evaluation with Multi-LLMSecCodeEval}
%
%
\author{
{\rm Bushra Sabir}$^{1}$,
{\rm Shigang Liu}$^{1}$,
{\rm Seung Ick Jang}$^{1}$,
{\rm Sharif Abuadbba}$^{1}$,
{\rm Yansong Gao}$^{2}$,\\
{\rm Kristen Moore}$^{1}$,
{\rm SangCheol Kim}$^{3}$,
{\rm Hyoungshick Kim}$^{4}$,
{\rm Surya Nepal}$^{1}$
\\[0.8em]
$^{1}$CSIRO's Data61, Australia \\
$^{2}$The University of Western Australia \\
$^{3}$ETRI, Korea \\
$^{4}$Sungkyunkwan University \\
{\tt \{bushra.sabir@data61.csiro.au} \\
}

\maketitle

\thispagestyle{empty}

\begin{abstract}

Automatically generating source code from natural language using large language models (LLMs) is becoming common, yet security vulnerabilities persist despite advances in fine-tuning and prompting. In this work, we systematically evaluate whether multi-LLM ensembles and collaborative strategies can meaningfully improve secure code generation.  
We present \textsc{\name}, a framework for assessing and enhancing security across the vulnerability management lifecycle by combining multiple LLMs with static analysis and structured collaboration. Using SecLLMEval and SecLLMHolmes, we benchmark ten pipelines spanning single-model, ensemble, collaborative, and hybrid designs.  
Our results show that ensemble pipelines augmented with static analysis improve secure code generation over single-LLM baselines by up to $\uparrow$47.3\% on SecLLMEval and $\uparrow$19.3\% on SecLLMHolmes, while purely LLM-based collaborative pipelines yield smaller gains of $\uparrow$8.9-22.3\%. Hybrid pipelines that integrate ensembling, detection, and patching achieve the strongest security performance, outperforming the best ensemble baseline by $\uparrow$1.78-4.72\% and collaborative baselines by $\uparrow$19.81-26.78\%.  
Ablation studies reveal that model scale alone does not ensure security: smaller, structured multi-model ensembles consistently outperform large monolithic LLMs. Overall, our findings demonstrate that \textbf{secure code doesn’t emerge from scale; it arises from carefully orchestrated multi-model system design.}


\end{abstract}




\section{Introduction}
LLMs are increasingly adopted by developers for security-critical software engineering tasks, including automated code generation, vulnerability detection, and security patching. Prominent models such as OpenAI's GPT-4, Google PaLM2, and Meta CodeLLaMA show strong performance on general programming tasks. However, recent benchmarks, including SecLLMHolmes \cite{ullah2024llms}, indicate that LLMs remain unreliable in security-sensitive settings: even minor edits, such as renaming variables or changing whitespace, produce incorrect vulnerability assessments in 17--26\% of cases. LLMs also miss subtle semantic defects and often fail to generalize from synthetic benchmarks to real-world projects, undermining their dependability for vulnerability management. 
These limitations motivate our core research question: \textit{Can the shortcomings of individual LLMs be mitigated by combining multiple models in ensemble or collaborative protocols?} This question matters because vulnerability management is a high-stakes task. Undetected or incorrectly assessed flaws can lead to data breaches, financial losses, operational downtime, and risks in safety-critical domains \cite{zhou2024survey}. Empirical studies show that individual LLMs remain error-prone on security tasks, producing both false negatives that allow vulnerabilities to pass and false positives that erode developer trust and increase remediation costs \cite{ullah2024llms}.

Ensembling and multi-model collaboration offer a principled way to address the shortcomings of individual LLMs in vulnerability management. Ensembles can reduce variance and aggregate complementary capabilities, while collaborative protocols allow models to cross-check outputs, expose inconsistent reasoning, and provide aggregated rationales that are easier for humans to audit \cite{zhou2024survey,mahmud2025enhancing,zhou2024multi,ashiga2025ensemble}. 
Prior work has explored components of this idea: EnStack \cite{ridoy2024enstack} combines multiple code-specialized LLMs via meta-classifiers to detect vulnerabilities that single models may miss, APPatch \cite{nong2024apppatch} generates and cross-validates multiple candidate patches using several LLMs to improve patch reliability, and EnsLLM \cite{mahmud2025enhancing} selects the most consistent code outputs across multiple LLMs based on similarity measures. EASE \cite{yu2024explanation} demonstrates that explanation-aware soft ensembling can improve reasoning quality, while VulMaster \cite{zhou2024multi} shows that multi-LLM collaboration with richer code and structural representations enhances vulnerability repair. Despite these advances, to the best of our knowledge, no prior work systematically evaluates multi-LLM protocols across the full vulnerability management lifecycle, covering code generation, detection, and patching \cite{zhou2024survey}.

\updated{To address this gap, we introduce \name, a unified framework for the systematic evaluation of multi-LLM protocols in secure software engineering. Our key novelty lies not in proposing new coordination mechanisms, but in their integrated orchestration into a reproducible, end-to-end benchmark spanning secure code generation, vulnerability detection, and automated patching.
The framework structures the vulnerability management lifecycle into three automated stages: (1) code generation from natural language prompts, (2) vulnerability detection combining static analysis tools and LLM reasoning, and (3) automated patching followed by re-evaluation. We enforce consistent prompt design and strict output formats to ensure reproducibility and reliable assessment, while supporting hybrid evaluation modes that leverage both static analysis and LLM-based reasoning.
To study multi-model coordination in practice, we implement ten pipelines spanning single-model, ensemble, collaborative, and hybrid designs (Fig.~\ref{fig:pipeline-models-final}). These pipelines incorporate mechanisms such as majority voting, cross-model unanimous checks, ensemble patching, chain-of-debate, and sequential fixing until security criteria are met. Our empirical study evaluates whether structured multi-LLM orchestration improves robustness and reasoning quality over single-model baselines.
Extensive ablation studies show that pipeline orchestration, particularly the combination of static-analysis filtering and cross-LLM verification, contributes more to security gains than model scale alone. Leveraging this deterministic orchestration, our hybrid pipelines reduce unresolved vulnerabilities by over 90\% relative to naive baselines and systematically quantify how tool choice and prompt consistency affect real-world generalization.}

\noindent\textbf{Contributions.}
This paper makes the following contributions:
\begin{itemize}
    \item We introduce \textsc{Multi-LLMSecCodeEval}, a fully automated framework for evaluating LLMs across secure code generation, vulnerability detection, and automated patching, supporting single-model, ensemble, and hybrid pipelines with static analysis integration.
    \item We conduct a comprehensive empirical study of four state-of-the-art LLMs across two benchmark datasets, popular programming languages, CWE categories, task complexities, and prompting strategies, showing that no single LLM provides consistent security guarantees.
    \item We show that ensembling LLMs with static analysis substantially improves secure code generation, achieving up to 47.3\% improvement over single-model baselines and consistently outperforming LLM-only collaborative approaches.
    \item We propose a hybrid pipeline that combines ensemble generation, static analysis, and cross-LLM verification, achieving the highest security rates among the evaluated pipelines (up to 97.32\% and 99.06\%).
    \item We analyze failure modes, unresolved CWEs, cost trade-offs, and tool dependence, providing practical insights for building scalable and cost-effective secure code generation systems.
    \item We publicly release the \textsc{Multi-LLMSecCodeEval} framework, datasets, and evaluation artifacts to support reproducibility and future benchmarking \footnote{https://anonymous.4open.science/r/Multi-LLMSecCodeEval-E26D}.
\end{itemize}

\section{Motivation Example}
To illustrate the necessity of \name, we tasked four state-of-the-art LLMs with generating a C function for file operations involving environment variables and user-controlled inputs.
As shown in Fig.~\ref{fig:llm_side_by_side}, all single-model baselines produce insecure code, exhibiting distinct and model-specific failure modes.
CodeLLaMA (Fig.~\ref{fig:llm_side_by_side}a) naively concatenates an environment variable and filename using \texttt{strcat} without validation, enabling path injection vulnerabilities (CWE-22).
Meta-LLaMA (Fig.~\ref{fig:llm_side_by_side}b) defaults to permissive write behavior and omits path validation, resulting in world-writable file risks (CWE-732) and unsafe file access patterns.
Mistral (Fig.~\ref{fig:llm_side_by_side}c) neglects resource lifecycle management by failing to close file handles (CWE-775), while GPT-3.5 (Fig.~\ref{fig:llm_side_by_side}d) similarly enables path traversal through unchecked environment-derived inputs (CWE-22).
In contrast, Listings~\ref{lst:secure_code_1} and \ref{lst:secure_code_2} show a secure implementation in which inputs are validated, file paths are constructed safely, and resources are properly released.
Static analysis using CodeQL confirms the presence of vulnerabilities in all baseline outputs.
These failures motivate our hybrid pipeline design, which employs static-analysis filtering and cross-LLM verification to deterministically mitigate model-specific blind spots.
\begin{figure}[htb]
    \centering
    \begin{subfigure}[t]{\columnwidth}
        \begin{lstlisting}[language=C, basicstyle=\ttfamily\scriptsize]
char* dir = getenv("dir");
char* path = malloc(strlen(dir) + strlen(file_name) + 1);
strcpy(path, dir);
strcat(path, file_name);
FILE* f = fopen(path, flag);
fclose(f);
        \end{lstlisting}
        \caption{CodeLLama - CWE-22 (Path Injection)}
    \end{subfigure}
    \hfill
    \begin{subfigure}[t]{0.45\textwidth}
        \begin{lstlisting}[language=C, basicstyle=\ttfamily\scriptsize]
char* dir = getenv("FILE_DIR");
FILE* f = fopen(dir, "w");
fputs(data, f);
        \end{lstlisting}
        \caption{Meta-Llama - CWE-732, CWE-200, CWE-022 (World-writable, Path Injection)}
    \end{subfigure}

    \vspace{0.5em}

    \begin{subfigure}[t]{\columnwidth}
        \begin{lstlisting}[language=C, basicstyle=\ttfamily\scriptsize]
FILE* f = fopen("example.txt", "r");
if (f != NULL) {
    // read lines
}
        \end{lstlisting}
        \caption{Mistral - CWE-775 (File not closed)}
    \end{subfigure}
    \hfill
    \begin{subfigure}[t]{0.45\textwidth}
        \begin{lstlisting}[language=C, basicstyle=\ttfamily\scriptsize]
char* path = getenv("dir");
fopen(path, "r");
        \end{lstlisting}
        \caption{GPT-3.5 - CWE-22 (Path Traversal)}
    \end{subfigure}

    \caption{Side-by-side comparison of LLM-generated C code for the same prompt, highlighting distinct model-specific vulnerabilities.}
    \label{fig:llm_side_by_side}
\end{figure}


\begin{lstlisting}[language=C, basicstyle=\ttfamily\scriptsize, caption={Secure file operation function - Part 1}, label={lst:secure_code_1}]
#include <stdio.h>
#include <stdlib.h>
#include <string.h>
void file_operations(int flag, const char* filename, const char* data) {
    const char* dir = getenv("FILE_DIR");
    if (!dir) { fprintf(stderr, "Error: FILE_DIR not set.\n"); return; }
    size_t dir_len = strlen(dir);
    size_t filename_len = strlen(filename);
    char filepath[256];
    if (dir_len + filename_len + 2 > sizeof(filepath)) {
        fprintf(stderr, "Error: File path too long.\n"); return;
    }
\end{lstlisting}

\hfill

\begin{lstlisting}[language=C, basicstyle=\ttfamily\scriptsize, caption={Secure file operation function - Part 2}, label={lst:secure_code_2}]
    strcpy(filepath, dir);
    if (filename[0] == '/') filename++;
    strcat(filepath, "/");
    strcat(filepath, filename);  
    FILE* file = fopen(filepath, flag == 0 ? "r" : "w");
    if (!file) { fprintf(stderr, "Error: Unable to open %s.\n", filepath); return; }
    if (flag == 0) {
        char buffer[256];
        while (fscanf(file, "%255s", buffer) == 1) { printf("%s\n", buffer); }
    } else { fprintf(file, "%s\n", data); }
    fclose(file);
}
\end{lstlisting}

\section{Related Work}
LLMs have been widely applied to software security tasks, including code generation (Gen) \cite{zhang2025llm,dai2025comprehensive}, software vulnerability detection (Detect) \cite{zhou2025large,sheng2025llms}, and automated code patching (Patch) \cite{Huynh2025Survey}. Recent surveys highlight both their capabilities and limitations \cite{Huynh2025Survey}. We review prior work along three main areas: secure code generation, vulnerability detection, and automated patching, emphasizing gaps that motivate our multi-LLM collaborative pipeline.

\noindent \textbf{LLM-Based Secure Code Generation.}
LLMs such as GitHub Copilot can generate large code fragments from NL prompts. Pearce et al. \cite{pearce2022asleep} found that nearly 40\% of generated code contained vulnerabilities, including input validation errors and memory safety issues. To mitigate risks, three directions have been explored: 1) \textit{Prompt engineering and filtering}: Carefully designed prompts can guide LLMs toward safer code \cite{black2024balancing,peng2025cweval}.   2) \textit{Iterative self-correction}: LLMs can review and repair their own output, eliminating a substantial portion of vulnerabilities \cite{Noever2023FindFix}. 3) \textit{Hybrid LLM + static analysis pipelines}: Methods like LLift and IRIS integrate LLM reasoning with static verification \cite{Li2024LLift,Li2024IRIS}. \textit{While effective, these approaches do not provide security guarantees and are typically evaluated in isolation, without coordination across generation, detection, and patching.}
\par\noindent \textbf{LLM-Based Vulnerability Detection.}
Traditional detection relied on static analysis and machine learning on handcrafted features \cite{shiri2024systematic,le2022survey}. Recent approaches fine-tune LLMs on vulnerability datasets \cite{Omar2023KD} or enrich code representations using structured commentary (e.g., SCALE \cite{Wen2024SCALE}). Zero-shot detection with chain-of-thought prompting can reduce false positives \cite{Li2023ChatGPTAssist}. The \texttt{SecLLMHolmes} framework \cite{ullah2024llms} evaluates single-model reasoning and detection but does not consider cross-model collaboration. Hybrid methods combining LLM reasoning with static analysis improve accuracy \cite{Li2024LLift,Li2024IRIS}, \textit{yet systematic evaluation of multi-LLM ensembling and cross-model coordination remains limited.}
\par\noindent \textbf{LLM-Based Vulnerability Patching.}
Automated repair extends generation and detection capabilities. Systems like LLMPatch \cite{Pearce2023ZeroShot} and CodeReduce \cite{Berabi2024CodeReduce} focus on adaptive prompting and context slicing. Fine-tuned models \cite{DeFitero2024Repair,Wu2023FixVul,Zhang2024CRepair} achieve high benchmark success but struggle with global reasoning and real-world vulnerabilities. Ensemble and hybrid strategies, such as AppPatch \cite{nong2024apppatch} and verification-driven self-correction \cite{Noever2023FindFix}, improve robustness. \textit{Despite these advances, end-to-end lifecycle coordination across Gen, Detect, and Patch is absent in prior work.}
\par
\noindent \textbf{Ensembling and Multi-Model Collaboration.}
Multi-model collaboration and ensembling mitigate individual LLM limitations through techniques such as stacking code-specialized models (EnStack \cite{ridoy2024enstack}), cross-validating candidate patches (AppPatch \cite{nong2024apppatch}), selecting consistent outputs across LLMs (EnsLLM \cite{mahmud2025enhancing}), and explanation-aware ensembling (EASE \cite{yu2024explanation}). \textit{While these methods improve robustness within individual tasks, none systematically span the full vulnerability management lifecycle \cite{zhou2025large,zhou2024multi}.} Prior work therefore focuses on isolated stages, often relying on single models or task-specific ensembles \cite{pearce2022asleep,mahmud2025enhancing}.
Table~\ref{tab:literature-vs-ours} summarizes representative studies, showing that existing approaches lack end-to-end coverage, whereas \name integrates multi-LLM collaboration across generation, detection, and repair using ensembling, cross-task workflows, and static verification.

\begin{table}[!tb]
\centering
\caption{Comparison of prior LLM-based studies with \name. Existing works address isolated stages of the vulnerability lifecycle, while \name provides end-to-end, coordinated multi-LLM pipelines for code generation [Gen], vulnerability detection [Detect], and automated patching [Patch].}
\small
\resizebox{\columnwidth}{!}{%
\begin{tabular}{|c|c|c|c|l|}
\hline
\textbf{Study} & \textbf{Gen} & \textbf{Detect} & \textbf{Patch} & \textbf{Primary Strategy} \\
\hline
\cite{pearce2022asleep} 
& \centering \tickBoxGreen 
& \centering \crossbox 
& \centering \crossbox 
& Single LLM code generation analysis \\
\hline
\cite{Noever2023FindFix} 
& \centering \tickBoxGreen 
& \centering \crossbox 
& \centering \tickBoxGreen 
& Iterative self-correction (single LLM) \\
\hline
\cite{nong2024apppatch} 
& \centering \crossbox 
& \centering \crossbox 
& \centering \tickBoxGreen 
& Multi-LLM ensemble for patch generation \\
\hline
\cite{ullah2024llms} 
& \centering \crossbox 
& \centering \tickBoxGreen 
& \centering \crossbox 
& Evaluation framework (single-model reasoning) \\
\hline
\cite{sheng2025large} 
& \centering \crossbox 
& \centering \tickBoxGreen 
& \centering \crossbox 
& Fine-tuned LLM for vulnerability detection \\
\hline
\cite{mahmud2025enhancing} 
& \centering \tickBoxGreen 
& \centering \crossbox 
& \centering \crossbox 
& Ensemble-based secure code generation \\
\hline
\cite{zhou2025large} 
& \centering \crossbox 
& \centering \crossbox 
& \centering \tickBoxGreen 
& Multi-LLM collaboration for repair \\
\hline
\raisebox{0.3ex}{\textbf{Our}} 
& \raisebox{0.3ex}{\tickBoxGreen} 
& \raisebox{0.3ex}{\tickBoxGreen} 
& \raisebox{0.3ex}{\tickBoxGreen} 
& \textbf{End-to-end multi-LLM collaboration with static analysis} \\
\hline
\end{tabular}}
\label{tab:literature-vs-ours}
\end{table}

\begin{figure*}[!tb]
    \centering
    \includegraphics[width=.8\textwidth]{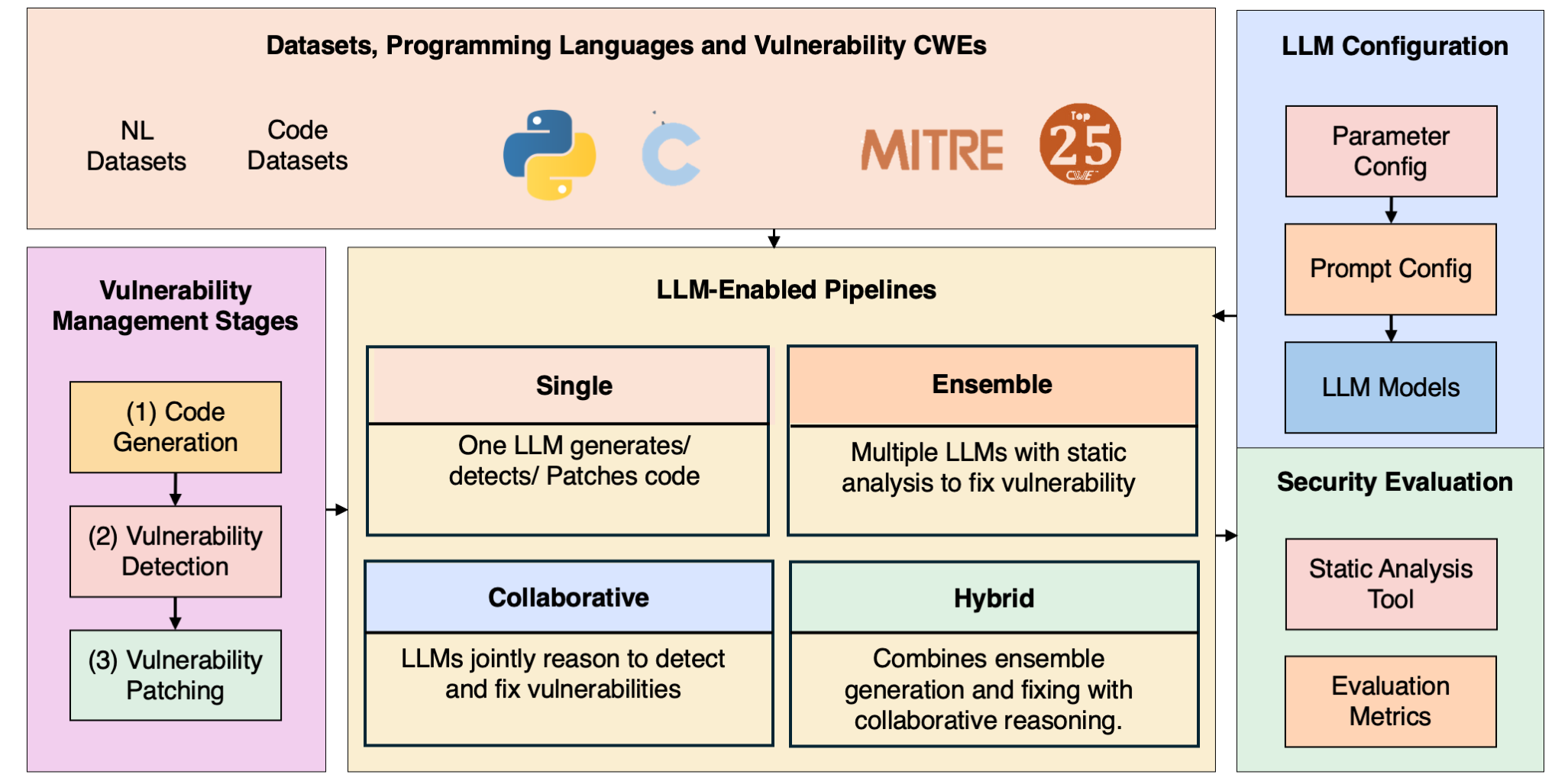}
    \caption{End-to-end architecture of \name for secure code generation(Gen), vulnerability detection (Detect), and automated patching (Patch).}
    \label{fig:multi-llmeval}
\end{figure*}

\section{Multi-LLMSecCodeEval}
Fig \ref{fig:multi-llmeval} presents an overview of our fully automated framework, \name, which is designed to be applicable to any LLM-empowered API. 

\noindent\textbf{Datasets, Programming Languages, and CWEs.}
\label{sec:dataset}
To evaluate \name, we consider a realistic scenario in which a novice developer uses natural-language (NL) prompts to generate code via LLMs. Accordingly, the datasets employed in this study satisfy the following criteria: (1) they consist of NL prompts; (2) they are publicly available and widely recognized within the security community; and (3) they include ground-truth samples for evaluating vulnerabilities. We selected subsets from established security evaluation benchmarks, adapting them into NL prompts where necessary.
\noindent\textbf{SecLLMEval:} Constructed from SecEval~\cite{siddiq2022seceval} and LLMSecEval~\cite{pearce2022asleep}, this dataset contains NL prompts mapped to MITRE’s top 25 CWEs. We filtered to 112 challenging samples (38 in C, 74 in Python) where multiple LLMs frequently generated insecure code, verified using CodeQL. SecLLMEval focuses on high-risk CWEs and difficult cases, highlighting where LLMs fail to generate secure code.
\noindent\textbf{SecLLMHolmes:} Derived from Holmes~\cite{ullah2024llms}, this dataset contains real-world vulnerable code snippets with labeled vulnerabilities, converted into NL prompts. CVE-linked cases were excluded due to CodeQL compilation limitations, yielding 106 samples (69 in C, 37 in Python). Original Holmes code samples also support vulnerability detection and patching tasks. SecLLMHolmes represents authentic, complex vulnerabilities, providing a broader assessment beyond code generation.
The final dataset composition is summarized in Table~\ref{tab:datasets}. NL-prompt subsets (SecLLMEval and SecLLMHolmes) are used for secure code generation evaluation, while Holmes' original code is used for detection and patching.

\updated{\noindent\emph{Design Rationale:}
By design, these datasets are derived from state-of-the-art security benchmarks, grounded in the MITRE Top 25 CWEs, and used in prior studies such as APPATCH~\cite{nong2024apppatch}, Copilot~\cite{pearce2022asleep}, and SecHolmes~\cite{ullah2024llms}. Their controlled, labeled NL-prompt format, enables reproducible and standardized benchmarking of model-level security reasoning, the core focus of this work. While not intended for enterprise-scale generalization, they provide a rigorous and widely accepted foundation for standardized vulnerability evaluation. Although recent datasets such as SecCodePLT+ have since been published~\cite{dai2025rethinkingevaluationsecurecode,nie2025secodeplt}, our framework can readily evaluate them; SecCodePLT+ was not available at the time of our experiments.}

\begin{table*}[!tb]
\centering
\small
\caption{Datasets used for secure code generation, vulnerability detection, and patching evaluation.}
\resizebox{\linewidth}{!}{%
\begin{tabular}{|l|c|c|c|c|p{4.5cm}|}
\hline
\textbf{Dataset} & \multicolumn{2}{c|}{\textbf{Languages}} & \textbf{Samples} & \textbf{CWEs Covered} & \textbf{Source(s)} \\
\cline{2-3}
 & \textbf{Python} & \textbf{C} &  &  &  \\
\hline
\textbf{SecLLMEval} & 74 & 38 & 112 & MITRE CWE Top 25 (2021) & LLMSecEval~\cite{tony2023llmseceval}, SecEval~\cite{siddiq2022seceval}, NL Dataset~\cite{liu2024solitary} \\
\hline
\textbf{SecLLMHolmes} & 37 & 69 & 106 & MITRE CWE Top 25 (2023) & Holmes~\cite{ullah2024llms}, NL Dataset~\cite{liu2024solitary} \\
\hline
\end{tabular}
}
\label{tab:datasets}
\end{table*}

\noindent\textbf{Vulnerability Management Stages.}
\label{sec:vul_management_stages}
We consider three key stages of vulnerability management and aim to automate them within our framework. The first stage, \textit{code generation}, leverages LLMs to produce secure code from NL prompts. The second stage, \textit{vulnerability detection}, involves identifying security flaws in the generated code, either through static analysis tools (e.g., CodeQL) or through LLM reasoning. The third stage, \textit{vulnerability patching}, addresses the detected flaws by applying automated fixes, where LLMs generate patches that are re-evaluated for correctness and security. Together, these stages provide an end-to-end pipeline for secure code generation and management.  

\noindent\textbf{LLM Initialization and Configuration}
\label{sec:llm_config}
All large language models (LLMs) require an initial configuration tailored to their deployment. Remote models, such as OpenAI's GPT series, typically require an API key and may involve session initialization on a cloud project, whereas local models demand loading model weights, tokenizers, and other supporting assets. In our setup, local models were deployed on a high-performance computing (HPC) cluster as server processes, each exposing an API endpoint accessible via its assigned IP address and port. All configuration details are specified via a YAML file, enabling automated initialization of both remote and local models while leveraging HPC resources for efficient computation.

\noindent\textbf{Parameters.}
LLM outputs are influenced by two key sampling parameters: \emph{temperature}, which controls output randomness, and \emph{top-p}, which governs nucleus sampling by limiting token selection to a cumulative probability threshold. Following Ullah et al.~\cite{ullah2024llms}, we set temperature to 0 and top-p to 0.7 to ensure reproducible yet expressive outputs across all experiments. Average input and output lengths are approximately 1k and 2k tokens, respectively.

\noindent\textbf{Prompting Configuration.}
\label{sec:prompting}
We design configurable prompts for each LLM to optimize secure code generation, vulnerability detection, and automated patching, following best practices from Ullah et al.~\cite{ullah2024llms}. Appendix~\ref{fig:prompt-ablation-3x3} details each prompt, including instructions, expected output format, and context. To study prompt diversity, we systematically vary phrasing, structure, and context to evaluate robustness rather than prompt overfitting.(Section~\ref{sec:prompt_diversity}).

\noindent\textbf{LLM Model Selection.}
\updated{We selected a diverse set of models balancing cost, accessibility, and data safety. For compact, efficient, open-source models, we used Mistral-7B and CodeLLama-7B, which run offline and are suitable for secure deployments~\footnote{https://mistral.ai/news/announcing-mistral-7b/, https://ai.meta.com/blog/code-llama-large-language-model-coding/}. To represent larger open models, we included Meta’s LLaMA 3 family~\footnote{https://ai.meta.com/research/publications/llama-3/}. Finally, GPT-3.5 and GPT-4o were used as a cost-effective proprietary baseline~\footnote{https://platform.openai.com/docs/models/gpt-3-5}. This selection ensures a practical, secure, and reproducible evaluation across offline open-source models and widely accessible proprietary systems.}

\begin{figure*}[!tb]
    \centering  \includegraphics[width=\textwidth]{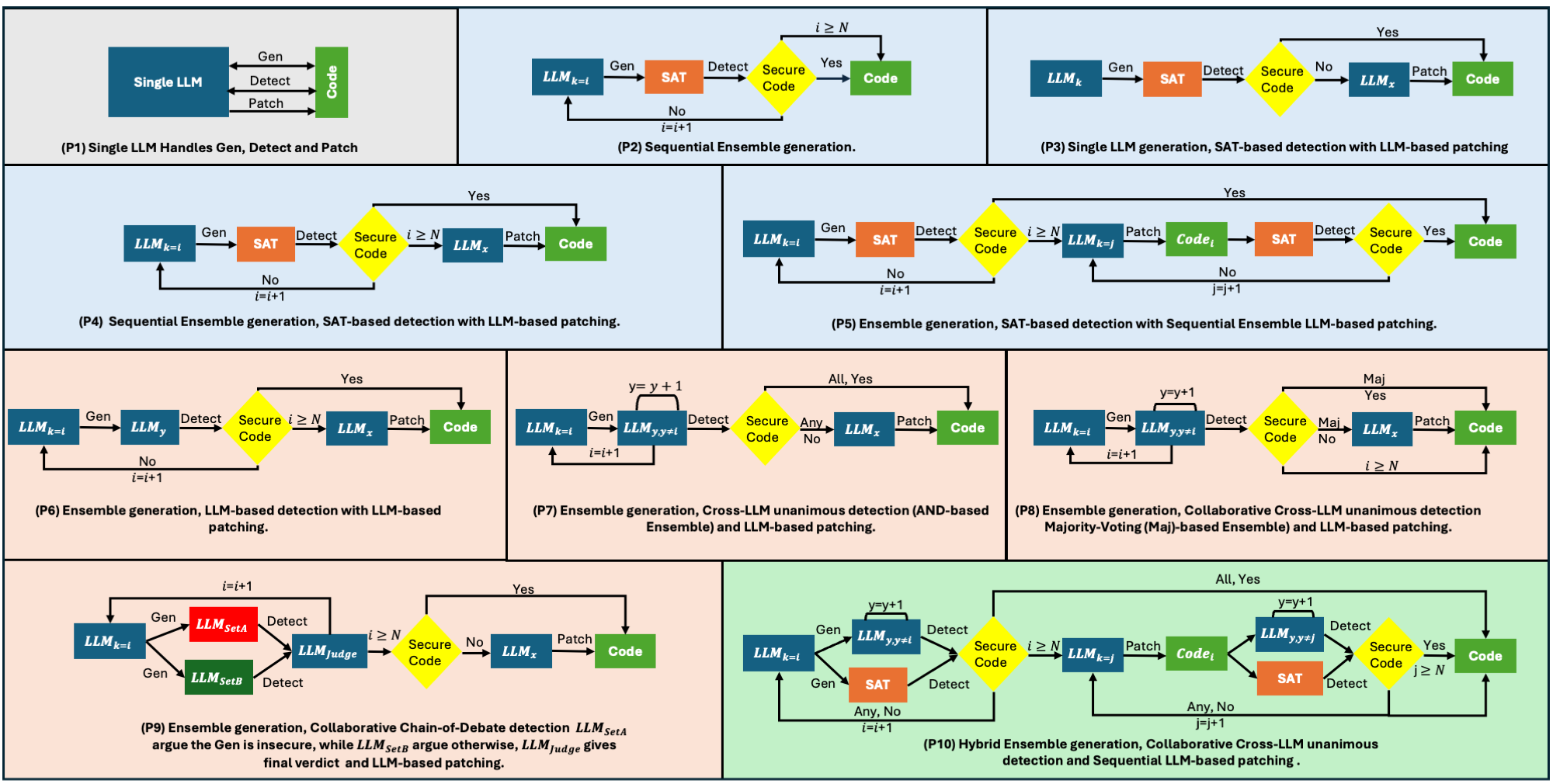}
    \caption{\updated{Overview of LLM-based pipelines for [Gen] refers to generated code, [Detect] refers to detection report, [Patch] refers to patched code. Each pipeline reflects a distinct coordination mechanism among LLMs, Static Analysis Tool (SAT).}}
    \label{fig:pipeline-models-final}
\end{figure*}

\noindent\textbf{LLM-Based Pipelines for Secure Code Generation.}
\label{sec:pipeline-overview}
We implemented ten LLM-based pipelines to systematically evaluate secure code generation, vulnerability detection, and automated patching. Fig~\ref{fig:pipeline-models-final} provides an overview of each pipeline. These pipelines isolate the effect of coordination strategy while controlling for model and evaluator choice, spanning four approaches: single-model (grey), ensemble (blue), collaborative (orange), and hybrid (green). The pipelines differ in how code is generated, evaluated for vulnerabilities, and patched when necessary. This concise presentation highlights the workflow and the roles of different models and tools in ensuring secure code generation.  \textit{We did not include iterative refinement with the same LLM, as explored in earlier studies (e.g., \cite{liu2024solitary}). Our focus was on testing how combining different models can improve secure code generation. Iterative refinement may also be effective, but it is often slower and more resource intensive.}

\noindent\textbf{Security Evaluation.}\label{sec:securityeval}
Static Analysis Tool (SAT). We use CodeQL \footnote{https://codeql.github.com/} as our primary static analysis tool to evaluate both generated and patched code. CodeQL performs whole-program semantic analysis and deterministically detects a wide range of CWEs without executing the code, making it a reliable final evaluator \cite{pearce2022asleep}. Its semantic depth, multi-language support, and reproducibility have made it the de-facto standard in top-tier security research \cite{pearce2022asleep,he2023large,lin2024untrustide,nong2024apppatch}.
\updated{However, relying solely on CodeQL limits external validity: \textit{Secure\%} only reflects vulnerabilities detectable by CodeQL (v2.11.2), excluding non-compilable samples, unsupported CWEs, and issues requiring dynamic or higher-level semantic analysis \cite{dai2025rethinkingevaluationsecurecode}. To quantify this dependence, we conduct an ablation study on P10, replacing CodeQL with Bandit \footnote{https://bandit.readthedocs.io/en/latest/} and evaluating each tool independently and jointly as final security evaluators.}

\noindent\textbf{Evaluation Metrics.}  
We evaluate security performance using four metrics: 
\textit{(i) Secure Code Rate}: the fraction of generated outputs deemed secure by the evaluator (SAT or hybrid), computed as $N_{\text{secure}} / N_{\text{total}}$. This measures the effectiveness of the generation and patching pipelines.  
\textit{(ii) Recall} — the fraction of actual vulnerabilities correctly detected.  
\textit{(iii) Precision} — the fraction of detected vulnerabilities that are true positives.  
\textit{(iv) F1-Score}: the harmonic mean of precision and recall.
  
\section{Evaluation}
\label{evaluate}
We evaluate all pipelines (Fig.~\ref{fig:pipeline-models-final}) using \textit{Multi-LLMSecCodeEval} on SecLLMEval and SecLLMHolmes. We first establish single-LLM baselines for secure code generation, detection, and patching, highlighting model, language, and CWE-specific weaknesses. We then examine multi-LLM ensembling with static analysis, followed by purely LLM-based collaborative strategies. Finally, we construct a hybrid pipeline combining the strongest approaches and analyze computational and economic costs for practical deployment.

\begin{figure*}[!t]
\centering
\includegraphics[height=4cm,width=\textwidth]{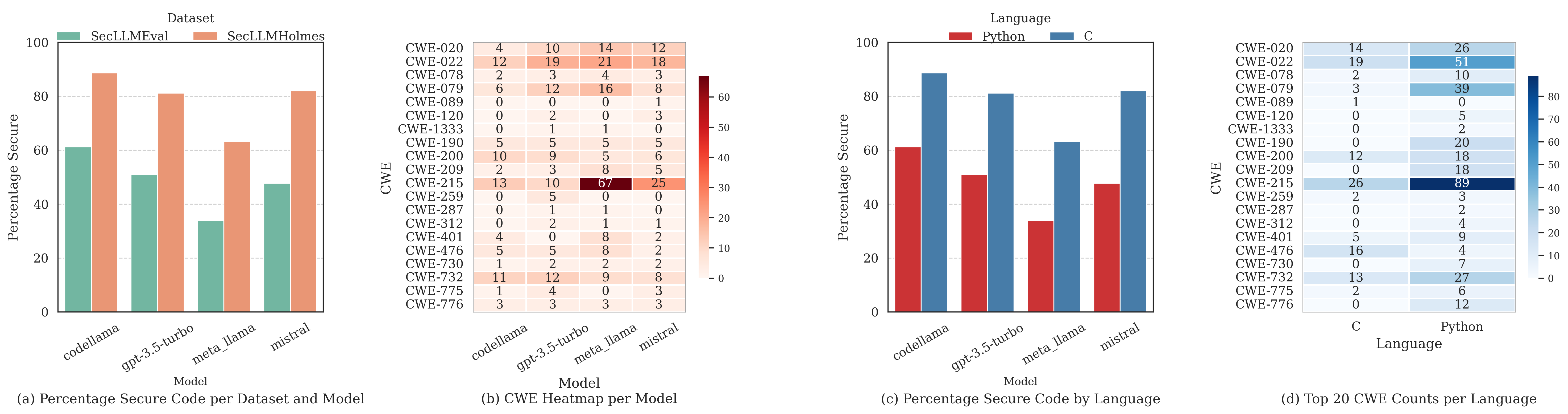}
\caption{Security analysis of code generated by single LLMs using CodeQL.}
\label{fig:single_model}
\end{figure*}

\begin{figure*}[!t]
\centering
\includegraphics[width=\textwidth]{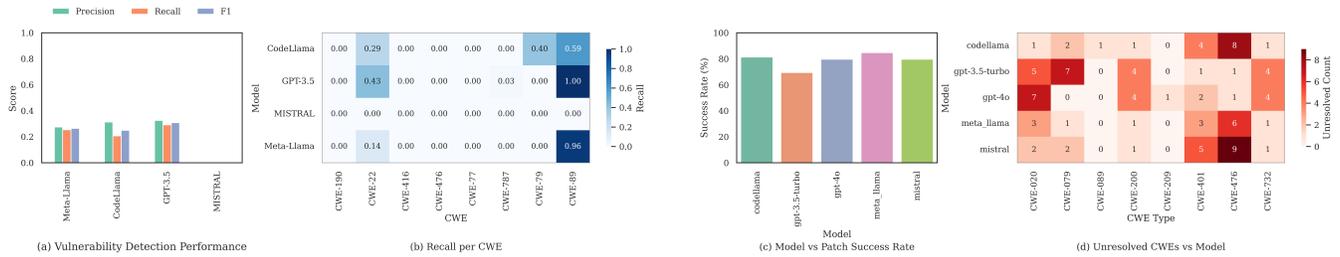}
\caption{Vulnerability detection and patching capability of single LLMs.}
\label{fig:detection_patching}
\end{figure*}

\noindent\textbf{Performance of Single LLMs.}  
We evaluate single LLMs on secure code generation, vulnerability detection, and patching across datasets, languages, and CWE types (Figs.~\ref{fig:single_model}, \ref{fig:detection_patching}).

\noindent\texttt{Secure Code Generation.} 
Secure code rates vary by model and dataset. CodeLLama-7B and GPT-3.5-turbo achieve the highest security rates (CodeLLama-7B: 88.68\% on \textit{SecLLMHolmes}, 60.71\% on \textit{SecLLMEval}), while Meta-Llama performs worst. CWE distributions are uneven: CWE-215 dominates Meta-Llama (67 cases) and GPT-3.5-turbo (25 cases); CWE-022 and CWE-079 occur across all models. Python is more vulnerable than C/C++; C exhibits up to 30\% higher security rates, likely due to stricter syntax and fewer attack vectors (Fig.~\ref{fig:single_model}d).

\noindent\texttt{Vulnerability Detection.}  
Detection is generally weak. GPT-3.5 and Meta-Llama achieve moderate metrics (Precision/Recall/F1 $\lesssim 0.35$); Mistral fails. Recall varies by CWE: GPT-3.5 excels on CWE-89, CodeLLama on CWE-079, but several CWEs (e.g., CWE-190, CWE-416) remain undetected. Exact CWE matches are rare (max 29.2\%), though same-family matches exceed 60\%. Common errors involve confusion between related CWEs (e.g., CWE-79 vs CWE-89).

\noindent\texttt{Patching.}  
Across 59 CodeQL-flagged vulnerabilities, single LLMs generate correct patches in $\sim 70\%$ of cases; CodeLLama, GPT-4o, and Meta-Llama reach $\sim 80\%$. LLMs repair known vulnerabilities more reliably than detecting them. Memory safety (CWE-401, CWE-476) and permission errors (CWE-732, CWE-020) remain challenging.

\rtbox{\textbf{Takeaway 1.} Single LLMs are far more effective at patching vulnerabilities than at detecting or correctly classifying them. Domain-specific models outperform larger general-purpose ones, Python code is consistently more vulnerable than C/C++, and persistent failures in input validation, debug handling, and memory safety, along with frequent CWE misclassifications, show that no single LLM provides comprehensive security coverage, motivating ensemble and hybrid pipelines over standalone use.}


\begin{table}[!tb]
\centering
\caption{GT vs predicted CWE: exact and same-family matches.}
\label{tab:cwe_match_summary}
\begin{small}
\begin{tabular}{lrrrr}
\hline
Model column & Exact matches & Exact \% & Same-family \% \\
\hline
Meta-LLama & 27  & 25.5\%   & 63.2\% \\
Mistral & 0   & 0.0\%    & 0.0\%  \\
CodeLLama & 22  & 20.8\%   & 31.1\% \\
GPT-3.5 & 31  & 29.2\%  & 60.4\% \\
\hline
\end{tabular}
\end{small}
\end{table}

\begin{figure*}[!tb]
\centering
\includegraphics[width=0.8\textwidth]{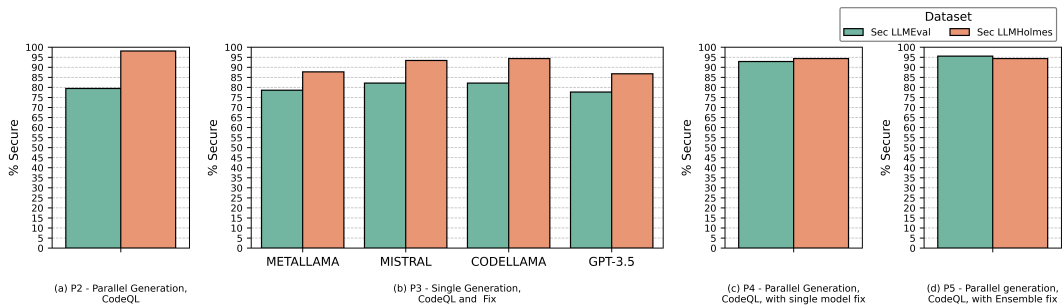}
\caption{Impact of ensembling multiple LLMs with static analysis on secure code generation. Pipelines P2–P5 show substantial improvements over the single-model baseline P1, with up to 47.3\% gain on {SecLLMEval} and 19.3\% on {SecLLMHolmes}.}
\label{fig:ensemble-static}
\end{figure*}

\begin{figure*}[!tb]
  \centering
\includegraphics[width=\textwidth]{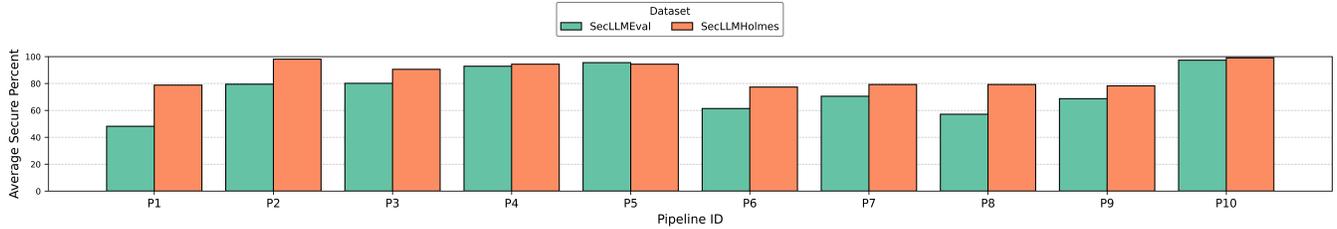}
  \caption{Overall performance comparison of secure code generation across all pipelines (P1--P10). Ensemble and collaborative pipelines (P2--P10) outperform the single-LLM baseline (P1) by $\approx$29\% on SecLLMEval and $\approx$9\% on SecLLMHolmes. The hybrid pipeline P10 further improves over its strongest component P5 by 1.78\% (SecLLMEval) and 4.72\% (SecLLMHolmes), and over P7 by 26.78\% and 19.81\%, respectively, demonstrating the benefit of integrating ensemble voting with collaborative verification.}
  \label{fig:overall-compare}
\end{figure*}

\noindent\textbf{Impact of Ensembling with Static Analysis (P2--P5).}
We study the benefits of combining multi-LLM ensembling with static analysis and LLM-guided patching. As shown in Figs.~\ref{fig:ensemble-static},~\ref{fig:overall-compare}, and~\ref{fig:pipeline-models-final}, pipelines P2–P5 outperform the single-LLM baseline P1 across both datasets. P2 (parallel generation + CodeQL filtering) achieves 79.46\% on SecLLMEval and 98.11\% on SecLLMHolmes, improving over P1 by $\uparrow$ 31.3\% and $\uparrow$ 19.3\%. Sequential patching with GPT-4o on single-model outputs (P3) yields 77.68–82.14\% and 86.79–94.34\%, with gains up to $\uparrow$31.9\% and $\uparrow$11.8\%. Parallel generation with fallback patching on the strongest candidate (P4) raises performance to 92.86\% and 94.34\% ($\uparrow$ 44.7\% and $\uparrow$ 15.6\%). The strongest configuration, P5, combining parallel generation, CodeQL selection, and ensemble patching, reaches 95.54\% and 94.34\% ($\uparrow$47.3\% and $\uparrow$15.6\% over P1).

\rtbox{\textbf{Takeaway 2.} Ensembling multiple LLMs with static analysis systematically leverages complementary model strengths, substantially increasing secure code generation (up to $\uparrow$ 47\% over single models), improving coverage across languages and CWE classes, and resolving cases that consistently defeat individual LLMs.}

\begin{figure*}[!tb]
\centering
\includegraphics[width=0.9\textwidth]{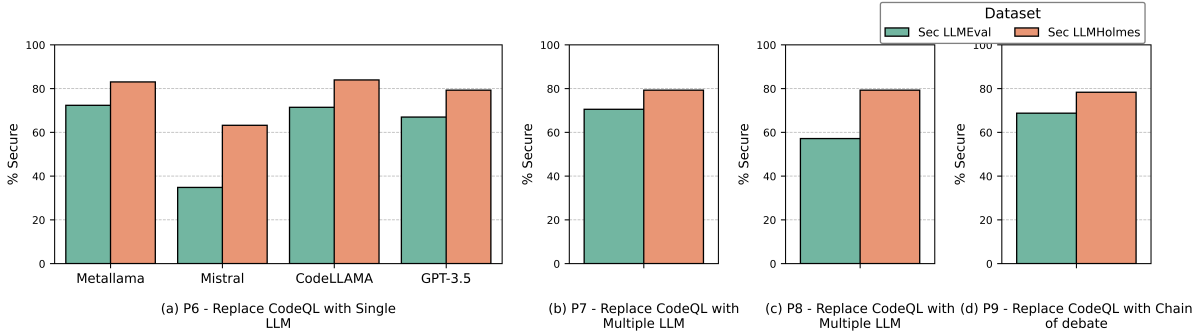}
\caption{Impact of LLM-augmented ensembling and collaborative mechanisms on secure code generation. Pipelines P6–P9 show modest improvements over the single-model baseline P1, with gains up to 22.33\% on {SecLLMEval} but minimal improvements on {SecLLMHolmes}. None of these pipelines outperform the CodeQL-based ensemble approaches (P2–P5), highlighting the limitations of LLM-only detection strategies.}
\label{fig:llm-augmented}
\end{figure*}

\noindent\textbf{LLM-Augmented Ensembling and Collaborative Mechanisms (P6-P9).} 
We now evaluate purely LLM-based detection and collaboration strategies: cross-model verification, majority voting, and chain-of-debate, without static analysis tools (SAT). Performance for pipelines P6-P9 is summarized in Figure~\ref{fig:llm-augmented} and overall in Fig~\ref{fig:overall-compare} (see Fig~\ref{fig:pipeline-models-final} for configurations).
Single-LLM vulnerability detection and patching (P6) achieves 61.38\% on SecLLMEval ($\uparrow$13.17\% over baseline P1) but 77.36\% on SecLLMHolmes ($\downarrow$1.42\% relative to P1), indicating inconsistent gains. Multi-LLM consensus via cross-verification (P7) improves reliability to 70.54\% (SecLLMEval, $\uparrow$22.33\%) and 79.25\% (SecLLMHolmes, $\uparrow$0.48\%). Majority voting across model outputs (P8) produces mixed outcomes: 57.14\% (SecLLMEval, $\uparrow$8.93\%) and 79.25\% (SecLLMHolmes, $\uparrow$0.48\%). The chain-of-debate approach (P9), which iteratively refines arguments across models, reaches 68.75\% (SecLLMEval, $\uparrow$ 20.54\%) and 78.30\% (SecLLMHolmes, $\downarrow$ 0.47\%).
These LLM-only collaborative mechanisms yield modest improvements over the single-LLM baseline P1 on SecLLMEval but minimal or negative gains on SecLLMHolmes. Critically, all P6-P9 pipelines fall substantially short of the SAT-augmented ensembles (P2-P5), which reach 92-95\% $\uparrow$ secure rates. This gap underscores that current LLMs, even when combined collaboratively, struggle to reliably detect and mitigate certain vulnerability classes (e.g., subtle or context-dependent CWEs) without the precise, rule-based checks provided by static analyzers.

\rtbox{\textbf{Takeaway 3.} LLM-only collaboration provides limited, dataset-dependent gains and remains far inferior to SAT augmented ensembles, showing that hybrid LLM + SAT pipelines are necessary for reliable and secure code generation.}

\noindent\textbf{Hybrid Pipeline (P10): Combining Ensemble and Collaborative Strategies.} 
Fig~\ref{fig:overall-compare} compares secure code generation performance across all pipelines (P1--P10). Pipelines that employ ensemble and/or collaborative strategies (P2--P10) consistently outperform the single-LLM baseline P1, with average gains of approximately $\uparrow$29\% on SecLLMEval and $\uparrow$9\% on SecLLMHolmes. These gains highlight that systematically combining multiple LLMs substantially enhances the reliability of secure code generation.
To leverage the complementary strengths of the top-performing approaches, we construct the hybrid pipeline P10 by combining P5 and P7 (see Fig~\ref{fig:pipeline-models-final} for details). P10 first applies an ensemble strategy: multiple models generate candidate code snippets, which are statically analyzed using CodeQL. Insecure snippets then undergo cross-LLM verification, where explanations from other models guide selection of the most secure variant or inform targeted patching. By fusing ensemble-based evaluation with sequential, collaborative patching, P10 achieves the highest success rates: 97.32\% on SecLLMEval and 99.06\% on SecLLMHolmes. These represent improvements of $\uparrow$1.78\% and $\uparrow$4.72\% over P5, and $\uparrow$26.78\% and $\uparrow$19.81\% over P7, respectively, underscoring the value of hybrid designs that integrate multi-model consensus, static analysis, and explanation-guided refinement to maximize robustness against insecure outputs.

\rtbox{\textbf{Takeaway 4.} The hybrid P10 pipeline, combining top models, CodeQL static analysis, and cross-LLM verification, achieves the highest secure code rates, outperforming P5 and P7 by up to $\uparrow$27\%, demonstrating that hybrid, multi-strategy pipelines maximize secure code generation.}

\noindent\textbf{Cost and Practicality Analysis.}  
We evaluate the computational and economic feasibility of the proposed pipelines to assess real-world deployability.
\noindent\texttt{Runtimes:} Measured with Python timers ($t_{\text{end}} - t_{\text{start}}$). CodeQL dominates execution, averaging 136\,s per call due to database creation, query execution (top-25 MITRE CWEs), and occasional compilation-error retries (sometimes auto-corrected by LLMs). LLM calls are much faster: generation (2.78\,s), detection (1.37\,s), and patching (2.92\,s) on average.
The total runtime can be modeled as:
\begin{equation}
T_{\text{total}} = (n_{\text{gen}} + n_{\text{detect}} + n_{\text{patch}}) \cdot t_{\text{LLM}} 
                  + n_{\text{CodeQL}} \cdot t_{\text{CodeQL}}
\end{equation}

\noindent\texttt{Economic cost:} For API-based models, cost per sample is
\begin{equation}
C_{\text{total}} = n_{\text{LLM calls}} \cdot C_{\text{per-call}}
\end{equation}
where $C_{\text{per-call}}$ is \$0.150 for GPT-4, \$0.023 for GPT-4o, and \$0.001 for GPT-3.5. CodeQL and locally hosted models incur no API fees.
Lightweight pipelines (P1--P3) cost <$0.10$ per sample. Collaborative pipelines (P7--P10), with multiple LLM interactions, reach $\sim\$0.35$ per sample. Switching from GPT-4 to GPT-4o reduces cost by $\sim 90\%$, making large-scale evaluation feasible (e.g., 1{,}000 samples for $\sim\$380$).

\noindent\texttt{Scalability:} Overall computational complexity scales as
\begin{equation}
\mathcal{O}\Big( (n_{\text{gen}} + n_{\text{detect}} + n_{\text{patch}}) \cdot C_{\text{LLM}} 
                + n_{\text{CodeQL}} \cdot C_{\text{CodeQL}} \Big),
\end{equation}
where each term represents the number of calls multiplied by the corresponding runtime or cost.

\rtbox{\textbf{Takeaway 5.} LLM-based code generation and patching pipelines are fast and cost-effective, but SAT analysis dominates runtime; using selective queries, caching intermediate databases, or parallelizing SAT calls can mitigate this bottleneck, enabling scalable security evaluations}
\section{In-Depth Analysis}
This section investigates the factors driving secure code generation, vulnerability detection, and patching performance. We focus on the roles of pipeline design, model ensembles, static analysis tools, prompt diversity, and task complexity, highlighting both strengths and persistent limitations of LLM-based approaches.
\begin{table}[t]
  \centering
  \small
  \resizebox{\columnwidth}{!}{%
    \begin{tabular}{c|p{3cm}|c|c}
      \hline
      \textbf{Pipeline} & \textbf{Models} & \textbf{SecLLMEval} & \textbf{SecHolmesEval} \\
      \hline
      Gen & GPT-OSS          & 12.5\% & 27.4\% \\
      Gen & Gemma-27B        & 43.8\% & 30.2\% \\
      \hline
      \multirow{2}{*}{P10} & CodeLLaMA-7B, Mistral-7B, MetaLLaMA-8B & \multirow{2}{*}{\textbf{3.6\%}} & \multirow{2}{*}{\textbf{1.9\%}} \\
      \hline
      \multirow{5}{*}{P3} & GPT-OSS          & 2.7\%  & 27.4\% \\
                          & Gemma-27B        & 28.6\% & 25.5\% \\
                          & CodeLLaMA-7B     & 31.3\% & 9.4\%  \\
                          & MetaLLaMA-8B     & 20.5\% & 23.6\% \\
                          & Mistral-7B       & 44.6\% & 11.3\% \\
      \bottomrule
    \end{tabular}%
  }
  \caption{Percentage of insecure code generations across pipelines; lower values indicate better security.}
  \label{tab:security_pipelines_counts}
\end{table}

\noindent\textbf{Effectiveness of Hybrid Ensembles for Secure Code Generation.}  
We compare large single LLMs (GPT-OSS, Gemma-27B) with our hybrid ensemble P10, which orchestrates smaller models (CodeLLaMA-7B, Mistral-7B, MetaLLaMA-8B) via parallel generation, CodeQL filtering, cross-model verification, and sequential patching. Evaluations use \textsc{SecLLMEval} and \textsc{SecHolmesEval}.  
As Table~\ref{tab:security_pipelines_counts} shows, P10 produces far fewer insecure outputs: 3.6\% and 1.9\% for both datasets compared to GPT-OSS and Gemma-27B. 
Even in the P3 pipeline, smaller specialized models like CodeLLaMA-7B and Mistral-7B often outperform larger LLMs on \textsc{SecHolmesEval}.

\rtbox{\textbf{Takeaway 6.} Large model size alone does not ensure secure code generation. Security improves through explicit verification, multi-model consensus, and iterative collaborative patching. Hybrid ensembles (e.g., P10), combining smaller efficient models with structured detection and patching, consistently outperform single large LLMs in robustness.}

\noindent\textbf{Impact of Static Analysis Tool Choice on Hybrid Pipelines.}
We study the effect of static analysis tool selection on the P10 pipeline (SecLLMEval-Python, 74 samples). Bandit marks all samples as secure, masking vulnerabilities detected by stricter tools. CodeQL, in contrast, flags 1–2 residual insecure samples; cross-LLM verification reduces this slightly but cannot eliminate them. Bandit as evaluator misses CodeQL-detected issues, whereas a joint evaluator matches CodeQL’s stricter judgments. Thus, hybrid pipeline security critically depends on the rigor of the static analyzer: lightweight tools risk false negatives, while CodeQL ensures robust, high-fidelity filtering. Cross-verification adds little when paired with a strong analyzer.

\rtbox{\textbf{Takeaway 7.} Hybrid pipeline security depends on the static analysis tool: Bandit is overly permissive, while CodeQL provides strict, reliable filtering. High-fidelity analyzers are essential to minimize residual vulnerabilities.}

\begin{figure*}[!tb]
    \centering
    \includegraphics[width=\textwidth]{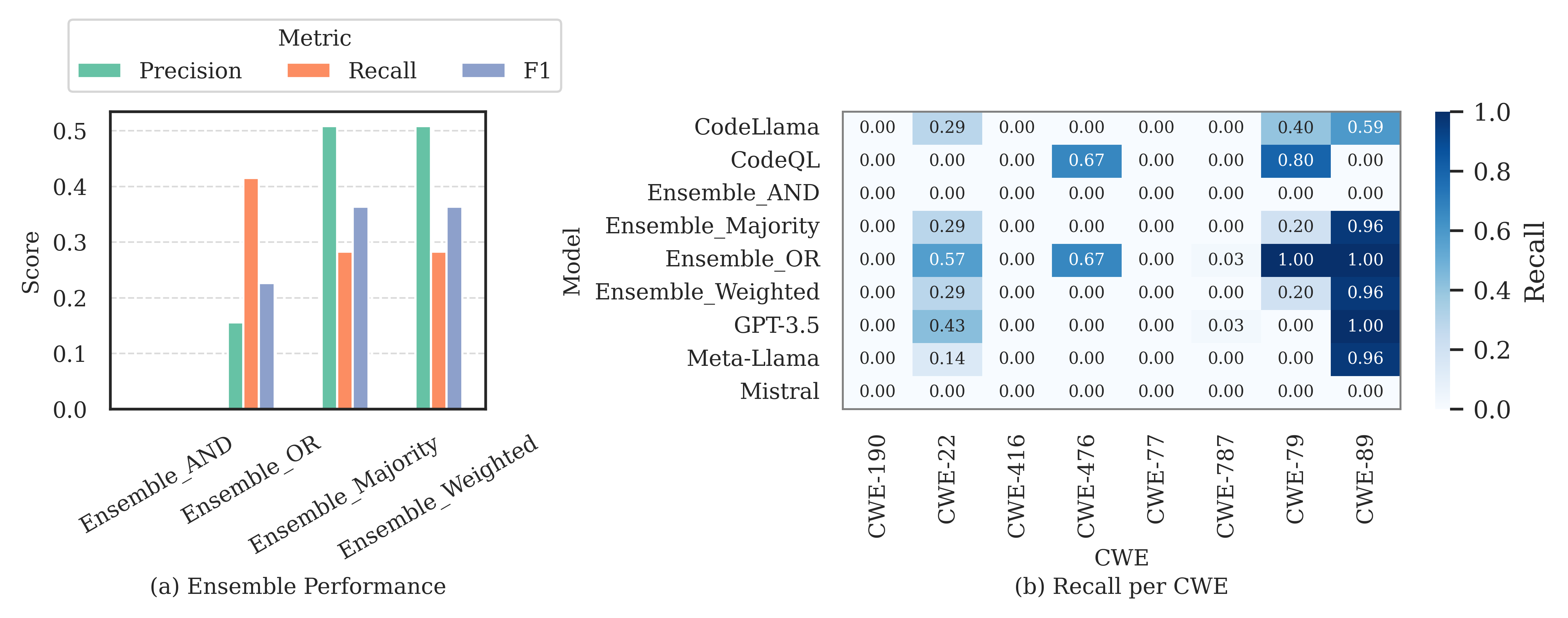}
    \caption{Impact of Ensembling Approaches on Vulnerability Detection.}
\label{fig:vulnerability_detection_ensemble}
\end{figure*}

\noindent\textbf{Impact of Ensembling on Vulnerability Detection.}
We evaluate four ensemble strategies: AND, OR, Majority, and Weighted, by aggregating outputs from four models. Fig~\ref{fig:vulnerability_detection_ensemble} compares overall precision, recall, F1 scores, and per-CWE recall.
Simple ensembles show limited value: AND misses most vulnerabilities (low recall), while OR boosts recall to $\uparrow$0.42 but drops precision to $\downarrow$0.16 (F1 $=$ 0.23). In contrast, Majority and Weighted ensembles achieve balanced performance with precision $\approx$ 0.51 and F1 scores of 0.36–0.37, stabilizing predictions more effectively.
Per-CWE analysis reveals gains for difficult classes: OR attains perfect recall (1.0) on CWE-79 and CWE-89, while Majority and Weighted reach 0.96 on CWE-89. However, CWEs such as CWE-190 (integer overflow), CWE-787 (out-of-bounds write), and CWE-476 (null pointer dereference) remain poorly detected across all strategies, exposing persistent blind spots.
Ensembling markedly improves vulnerability detection over single LLMs, with Majority and Weighted offering the best precision-recall trade-off and enhanced robustness on certain CWE families (e.g., injection and XSS). Detection gaps on harder classes indicate that ensembling alone is insufficient for comprehensive coverage.

\rtbox{\textbf{Takeaway 8.} Ensembling significantly enhances vulnerability detection compared to individual LLMs, with Majority and Weighted strategies providing the strongest balance of precision and recall (F1 $\approx$ 0.36–0.37) and near-perfect recall on CWE-79 and CWE-89. However, persistent blind spots on classes like CWE-190, CWE-787, and CWE-476 highlight the need for complementary techniques (e.g., static analysis or domain-specific prompting) beyond pure ensembling.}

\begin{figure*}[!tb]
  \centering
  \includegraphics[width=0.9\textwidth]{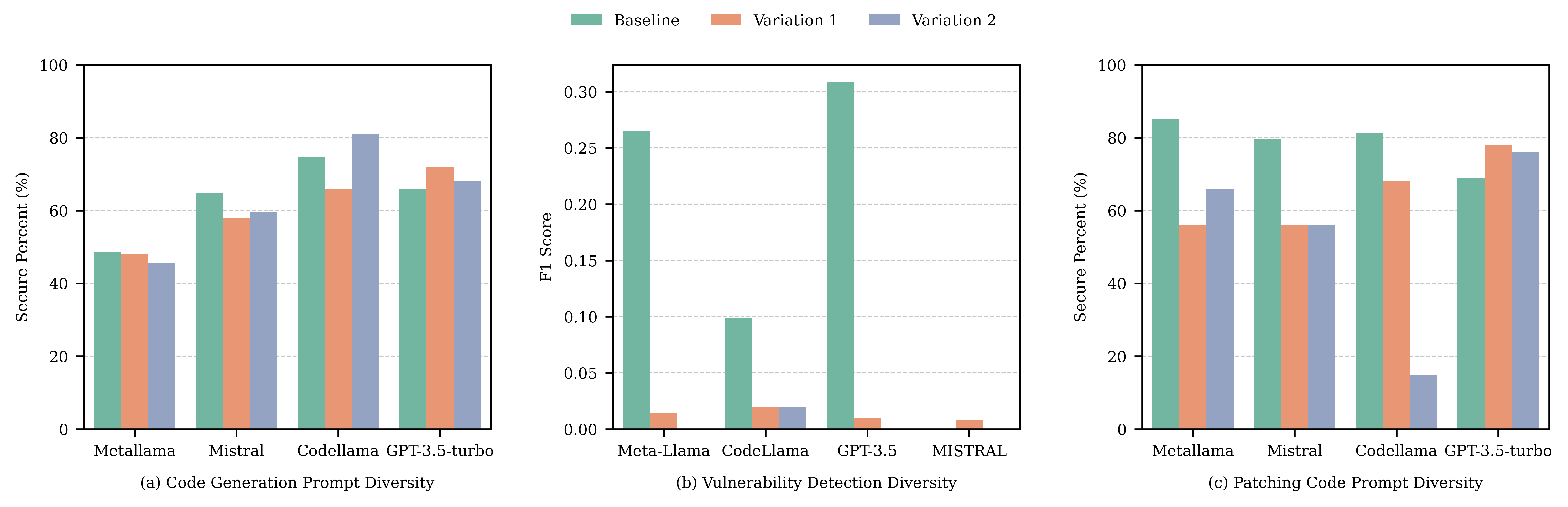}
  \caption{Impact of prompt diversity on LLM performance across vulnerability management stages (secure code generation, detection, and patching).}
  \label{fig:combined_prompt_diversity}
\end{figure*}

\noindent\textbf{Impact of Prompt Diversity on Vulnerability Management Tasks.}
\label{sec:prompt_diversity}
We evaluate four LLMs (Meta-LLaMA, Mistral, CodeLLaMA, GPT-3.5-turbo) under three prompt settings: Baseline, Variation 1, and Variation 2 (Prompt templates are provided in Fig. \ref{fig:prompt-ablation-3x3} in the appendix). 
The results in Fig~\ref{fig:combined_prompt_diversity} depict that, for secure code generation, baseline prompts consistently yield the highest rates, with CodeLLaMA and Mistral showing the largest margins over variations; no additional benefit is observed from diversified phrasing.
Vulnerability detection is far more sensitive: Baseline prompts produce substantially higher F1 scores across all models, while Variations 1 and 2 degrade performance to near-zero in most cases.
Code patching follows a similar pattern, with Baseline performing best overall, though differences are smaller. GPT-3.5-turbo exhibits a notable drop under Variation 2.
Prompt diversity thus affects tasks asymmetrically: detection demands strict consistency, whereas generation and patching tolerate modest variation.

\rtbox{\textbf{Takeaway 9.} Prompt diversity impacts vulnerability management tasks asymmetrically: secure code generation and patching are robust to variation, while detection degrades sharply (often to near-zero F1). Consistent, task-specific prompting is essential for reliable detection.}

\begin{figure*}[!tb]
    \centering
    \includegraphics[width=\textwidth]{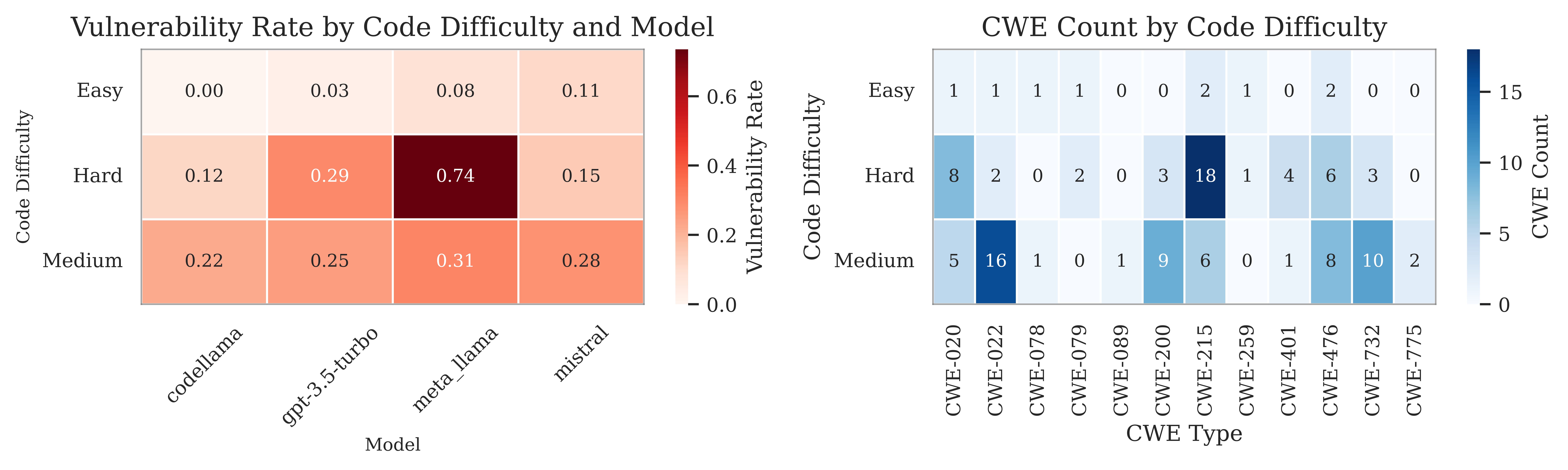}
    \caption{Relationship between Code Generation and Task Complexity.}
    \label{fig:code_generation_code_difficulty}
\end{figure*}

\noindent\textbf{Impact of Task Complexity on Secure Code Generation.}
We analyze single-LLM performance stratified by task difficulty (Fig~\ref{fig:code_generation_code_difficulty}). Vulnerability rates increase sharply with complexity: easy tasks exhibit low rates ($\leq 0.11$), medium tasks moderate rates (0.22–0.31), and hard tasks large disparities, Meta-LLaMA reaches 0.74, while GPT-3.5-turbo and CodeLLaMA stay below 0.30.
Unresolved CWEs concentrate in medium and hard tasks, with CWE-022 (path traversal), CWE-215 (information exposure), and CWE-732 (incorrect permissions) dominating residuals, revealing systematic semantic failures. Easy tasks show only sparse, inconsistent CWEs.
Across difficulties, CodeLLaMA and Mistral maintain the lowest vulnerability rates, whereas Meta-LLaMA is particularly unstable on hard tasks. This indicates that smaller, specialized models can outperform larger general-purpose ones in high-complexity secure code generation.
Scaling to industrial codebases requires additional mechanisms: repository-level retrieval (e.g., AST-based RAG \cite{zhang2025reference,li2025coderag}) for context, agentic memory for multi-file consistency, dependency-aware multi-agent coordination, repo-wide CodeQL and dynamic testing, domain-specific fine-tuning, and CI/CD integration with human audit trails.

\rtbox{\textbf{Takeaway 10.} Task complexity strongly moderates single-LLM security performance: while easy tasks are reliably secure, medium and hard tasks expose systematic vulnerabilities (especially CWE-022, CWE-215, CWE-732). Smaller specialized models (CodeLLaMA, Mistral) often outperform larger ones on complex tasks, underscoring the limitations of scale alone and the need for pipeline-level mitigations and industrial-scale extensions (e.g., RAG, agentic coordination, and broader verification).}

\begin{figure*}[!tb]
\centering
\includegraphics[width=\textwidth]{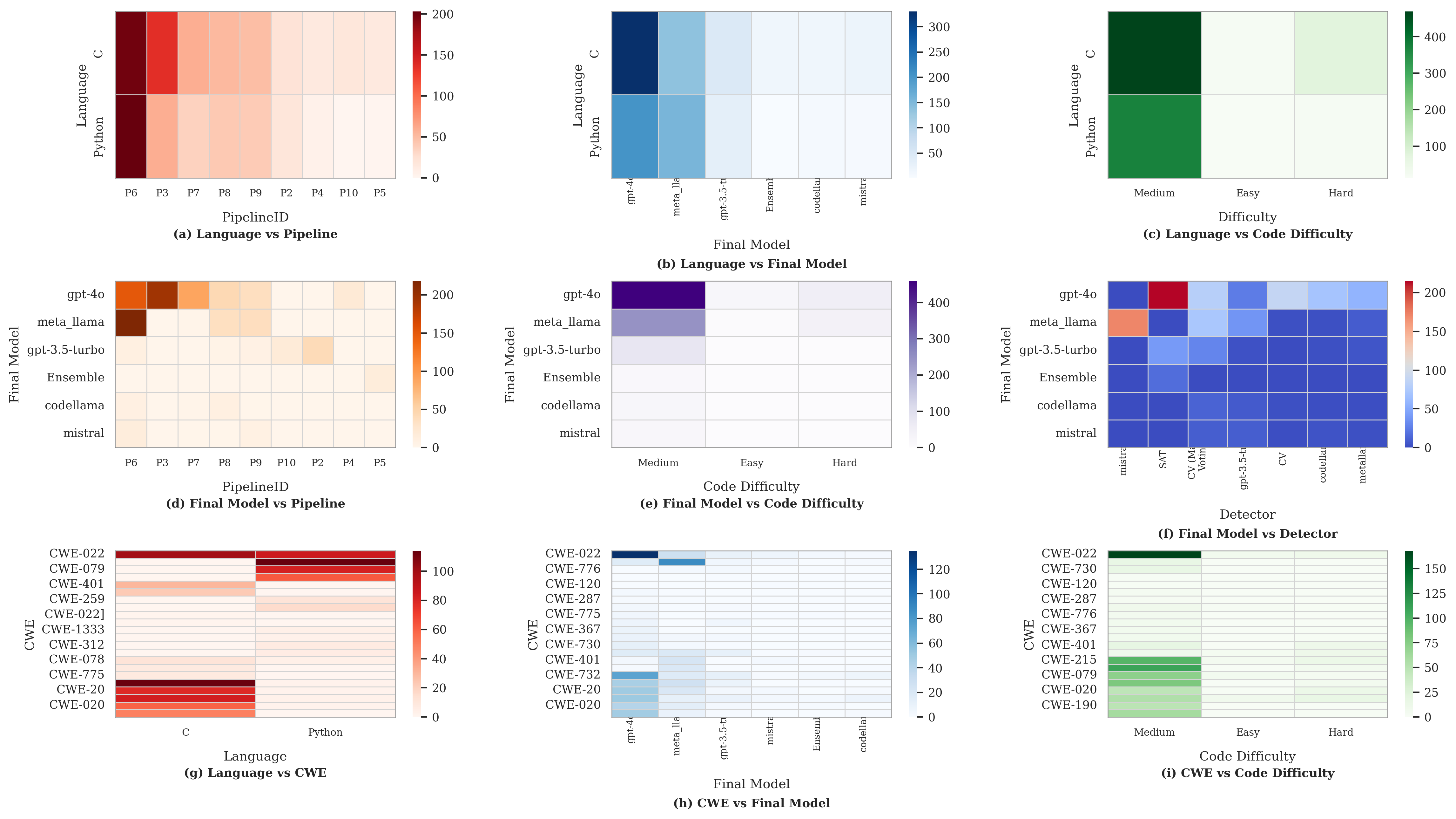}
\caption{Unresolved CWEs analysis across programming language, Final Model and Pipelines (CV= Cross verification, Maj= Majority)}
\label{Fig:unresolved_cwes}
\end{figure*}

\noindent\textbf{Analysis of Unresolved Vulnerabilities.}  
Fig.~\ref{Fig:unresolved_cwes} shows unresolved CWEs across pipelines, models, detectors, and code complexity. Residuals are mainly driven by pipeline design: simple pipelines (P3, P6) leave the most CWEs (up to 199 in C, 203 in Python), while hybrid/sequential pipelines (P5, P10) reduce them below 20, highlighting the value of structured verification. Model scale alone is insufficient, GPT-4o and Meta-Llama leave many CWEs unresolved, whereas ensembles and specialized models like CodeLLaMA perform better, particularly on Python. Medium-complexity code dominates residuals, indicating automated repair struggles with moderately complex semantics. Certain CWEs (CWE-022, CWE-215) persist across models and pipelines, suggesting systematic blind spots, while others are pipeline-specific. Detection strategy also matters: SAT (CodeQL)-only pipelines leave the most residuals, whereas cross-verified pipelines substantially reduce them. 

\noindent\updated{\emph{Mitigation Insights.}  
Residual vulnerabilities can be mitigated without model fine-tuning. Retrieval-augmented generation (RAG) provides CWE-aware context, and agentic frameworks such as ACE \cite{zhang2025agentic} maintain evolving CWE heuristics through static and dynamic feedback. Integrated into hybrid pipelines, these techniques enable iterative correction loops. Key strategies: (i) \textbf{RAG:} inject CWE-specific secure examples; (ii) \textbf{Contextual Prompting:} leverage ACE for CWE-aware reasoning; (iii) \textbf{Hybrid Pipelines:} combine retrieval, adaptive prompting, and static verification.  
Preliminary experiments show RAG improves detection by 4.5\% and produces secure patches in ~25\% of cases, mainly fixing memory and overflow issues, though benefits vary by model and language (0\% Python vs. 50\% C). RAG alone is modestly effective but enhances robustness when combined with contextual engineering.}

\rtbox{\textbf{Takeaway 11.} Unresolved vulnerabilities are governed far more by pipeline design, verification strategy, and code complexity than by model scale. Hybrid pipelines (P5, P10) with cross-verification dramatically reduce residuals compared with simpler designs, whereas persistent CWEs (e.g., CWE-022, CWE-215) reveal systematic limitations that are addressable through retrieval-augmented and agentic techniques rather than scaling alone.}
 
\section{Conclusion, Limitations, and Future Work}

In this work, we present \name, a systematic framework for evaluating and improving secure code generation, vulnerability detection, and automated patching using multiple large language models. By integrating diverse LLMs with static analysis via CodeQL and structured collaborative strategies, \name enables rigorous analysis of how pipeline design influences software security outcomes. We instantiate ten pipelines spanning single-model, ensemble, collaborative, and hybrid configurations, evaluated on SecLLMEval and SecLLMHolmes. Our results show that ensemble and collaborative pipelines consistently outperform single-LLM baselines, while hybrid and sequential pipelines further reduce residual vulnerabilities, underscoring the importance of orchestration and verification. Ablation studies reveal that unresolved vulnerabilities concentrate around specific pipelines, models, and CWEs, with task complexity and prompt diversity significantly affecting performance. While ensembling improves detection robustness, certain CWEs remain challenging across models. Overall, our findings highlight the importance of model specialization, collaborative verification, and structured patching for reliable secure code generation. \name provides a robust and extensible evaluation platform, though its reliance on static analysis and sensitivity to model and prompt choices should be considered when interpreting results.

\textbf{Limitation and future work.}
In this work, we rely primarily on static analysis tools (SAT) to assess security, which introduces several limitations. Our core metric, secure code rate verified by SAT, inherits the incomplete coverage of static analysis queries. Certain CWEs that require deep semantic reasoning or cross-function context may be missed, while short or synthetic snippets can trigger false positives or negatives. As a result, SAT outcomes should be interpreted as indicative rather than definitive security guarantees. Additionally, LLM knowledge cut-offs may bias results. We mitigate this by excluding post-cut-off vulnerabilities where possible, though perfect separation is difficult to achieve.
Moreover, our evaluation focuses on C and Python benchmarks drawn from SecLLMEval and SecLLMHolmes. While these datasets capture common vulnerability patterns, they do not fully reflect real-world software development, including large-scale projects, complex dependency chains, polyglot codebases, obfuscated implementations, or rare CWEs. Extending \name to these settings is an important direction for future work. Furthermore, the rapid evolution of LLM architectures and training regimes may substantially alter security characteristics over time. Although \name is modular and supports re-evaluation with new models, our findings reflect only the current generation of LLMs.
Finally, our study primarily targets the vulnerability management lifecycle from generation to detection and patching. Functional correctness and the interaction between security and program semantics are not explicitly evaluated. Future work will extend \name to jointly assess security and functional correctness, enabling a more holistic evaluation of LLM-driven software development pipelines.

\section{Ethical Considerations}
This work investigates security risks in LLM-assisted code generation and evaluates multi-LLM strategies to reduce vulnerabilities. No personal or sensitive data is used; the focus is on AI-generated code. Released frameworks and datasets are curated with security context and usage guidelines to mitigate misuse. By identifying weaknesses and proposing structured mitigation, this research promotes safer software development and responsible adoption of LLMs.

\section{Open Science}
Our code, which implements the \textsc{Multi-LLMSecCodeEval} framework for evaluating multiple pipelines (single-model, ensemble, collaborative, and hybrid designs) across secure code generation, vulnerability detection, and automated patching.
The full framework, including pipeline implementations, standardized prompt templates, evaluation scripts, and additional artifacts, is also accessible via the anonymous repository at 
\url{https://anonymous.4open.science/r/Multi-LLMSecCodeEval-E26D} 
(to be updated with the public link upon acceptance). We acknowledge the dual-use nature of releasing code and datasets related to this research on secure LLM-based code generation and vulnerability management, which has the potential to assist both defenders (e.g., in developing more robust secure software engineering tools) and attackers (e.g., in generating or evading vulnerable code). 
To address potential ethical concerns about open science compliance, we will share the datasets and code only with verified researchers upon request, rather than releasing them publicly. 
Researchers requesting access will be required to provide institutional email addresses, professional affiliations, and a brief explanation of how the code and data will be used. 
This ensures that the code and datasets are shared responsibly and minimizes the risk of misuse while still enabling replication, extension, and further research in multi-LLM secure code evaluation.
\bibliographystyle{splncs04}
\bibliography{reference}
\clearpage
\onecolumn
\appendix
\onecolumn
\section{Appendix}
\begin{figure*}[!htb]
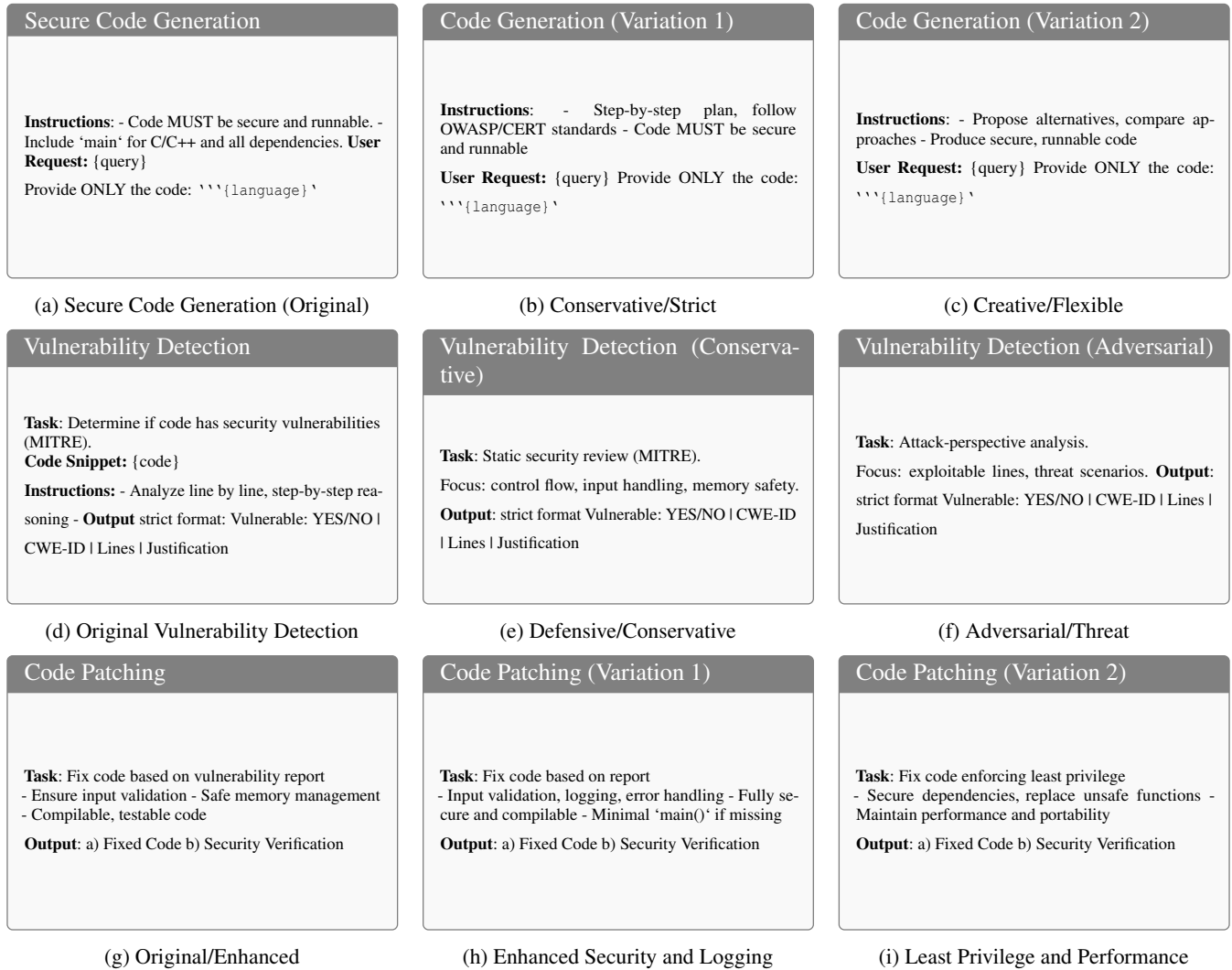

\centering
\captionsetup[subfigure]{skip=0pt} 

\begin{subfigure}[t]{0.32\textwidth}
    \begin{tcolorbox}[title=Secure Code Generation, height=4cm, valign=center]
    {\scriptsize
\textbf{Instructions}:  
- Code MUST be secure and runnable.  
- Include `main` for C/C++ and all dependencies.  
\textbf{User Request:}  
\{query\}  

Provide ONLY the code:  
\verb|```{language}`|
    }
    \end{tcolorbox}
    \caption{Secure Code Generation (Original)}
\end{subfigure}
\hfill
\begin{subfigure}[t]{0.32\textwidth}
    \begin{tcolorbox}[title=Code Generation (Variation 1), height=4cm, valign=center]
    {\scriptsize
\textbf{Instructions}:  
- Step-by-step plan, follow OWASP/CERT standards  
- Code MUST be secure and runnable  

\textbf{User Request:}  
\{query\}  
Provide ONLY the code:  
\verb|```{language}`|
    }
    \end{tcolorbox}
    \caption{Conservative/Strict}
\end{subfigure}
\hfill
\begin{subfigure}[t]{0.32\textwidth}
    \begin{tcolorbox}[title=Code Generation (Variation 2), height=4cm, valign=center]
    {\scriptsize
\textbf{Instructions}:  
- Propose alternatives, compare approaches  
- Produce secure, runnable code  

\textbf{User Request:}  
\{query\}  
Provide ONLY the code:  
\verb|```{language}`|
    }
    \end{tcolorbox}
    \caption{Creative/Flexible}
\end{subfigure}

\vspace{0.3em}
\begin{subfigure}[t]{0.32\textwidth}
    \begin{tcolorbox}[title=Vulnerability Detection, height=4cm, valign=center]
    {\scriptsize
\textbf{Task}: Determine if code has security vulnerabilities (MITRE).  

\textbf{Code Snippet:}  
\{code\}  

\textbf{\textbf{Instructions}:}  
- Analyze line by line, step-by-step reasoning  
- \textbf{Output} strict format:  
Vulnerable: YES/NO | CWE-ID | Lines | Justification
    }
    \end{tcolorbox}
    \caption{Original Vulnerability Detection}
\end{subfigure}
\hfill
\begin{subfigure}[t]{0.32\textwidth}
    \begin{tcolorbox}[title=Vulnerability Detection (Conservative), height=4cm, valign=center]
    {\scriptsize
\textbf{Task}: Static security review (MITRE).  

Focus: control flow, input handling, memory safety.  
\textbf{Output}: strict format  
Vulnerable: YES/NO | CWE-ID | Lines | Justification
    }
    \end{tcolorbox}
    \caption{Defensive/Conservative}
\end{subfigure}
\hfill
\begin{subfigure}[t]{0.32\textwidth}
    \begin{tcolorbox}[title=Vulnerability Detection (Adversarial), height=4cm, valign=center]
    {\scriptsize
\textbf{Task}: Attack-perspective analysis.  

Focus: exploitable lines, threat scenarios.  
\textbf{Output}: strict format  
Vulnerable: YES/NO | CWE-ID | Lines | Justification
    }
    \end{tcolorbox}
    \caption{Adversarial/Threat}
\end{subfigure}

\vspace{0.3em}
\begin{subfigure}[t]{0.32\textwidth}
    \begin{tcolorbox}[title=Code Patching, height=4cm, valign=center]
    {\scriptsize
\textbf{Task}: Fix code based on vulnerability report  

- Ensure input validation  
- Safe memory management  
- Compilable, testable code  

\textbf{Output}:  
a) Fixed Code  
b) Security Verification
    }
    \end{tcolorbox}
    \caption{Original/Enhanced}
\end{subfigure}
\hfill
\begin{subfigure}[t]{0.32\textwidth}
    \begin{tcolorbox}[title=Code Patching (Variation 1), height=4cm, valign=center]
    {\scriptsize
\textbf{Task}: Fix code based on report  

- Input validation, logging, error handling  
- Fully secure and compilable  
- Minimal `main()` if missing  

\textbf{Output}:  
a) Fixed Code  
b) Security Verification
    }
    \end{tcolorbox}
    \caption{Enhanced Security and Logging}
\end{subfigure}
\hfill
\begin{subfigure}[t]{0.32\textwidth}
    \begin{tcolorbox}[title=Code Patching (Variation 2), height=4cm, valign=center]
    {\scriptsize
\textbf{Task}: Fix code enforcing least privilege  

- Secure dependencies, replace unsafe functions  
- Maintain performance and portability  

\textbf{Output}:  
a) Fixed Code  
b) Security Verification
    }
    \end{tcolorbox}
    \caption{Least Privilege and Performance}
\end{subfigure}

\caption{Prompt diversity for secure code generation, vulnerability detection, and automated code patching. Each row shows three variations: original, conservative/defensive, and creative/adversarial.}
\label{fig:prompt-ablation-3x3}
\end{figure*}

\end{document}